# The Broken Engine of Szilard

Quanmin Guo

*School of Physics and Astronomy, University of Birmingham, Birmingham B15 2TT, UK*

**Abstract**

A crucial step in the operation of the Szilard engine is the isothermal expansion of a single particle system. This expansion, seemingly a natural consequence, is unable to proceed when the piston is considered appropriately to be a thermal system subject to fluctuations in energy at a constant temperature. The Szilard engine is thus unable to complete its cycle and the effort to locate the mysterious missing entropy of $k\ln 2$ turns out to be superfluous.

Recently, a number of Szilard-engine-type of devices have been constructed to demonstrate the relationship between information and thermodynamics [1-6]. This has stimulated a great deal of interest in theoretical modeling of such devices [7-9]. The original Szilard engine introduced by Leo Szilard in 1929 [10] consists of a single molecule confined in a box which is in thermal contact with a large thermal reservoir, Fig. 1. By inserting a piston, the volume of the box is divided into two-halves. The single molecule ends up in either the left or the right hand side of the piston. The pressure from this single molecule pushes the piston towards one end of the box. If the piston is coupled to some mechanical device, then work can be extracted due the isothermal expansion of this one-molecule engine. Once the piston reaches to the end of the box, it is removed and the system returns to its starting point. Because the whole system is in thermal contact with the reservoir at temperature T, the work done by the expanding piston is $kT \ln 2$. This amount of work is provided by heat transfer from the reservoir to the engine. Hence over one complete cycle, the reservoir loses $kT \ln 2$ in heat corresponding to a net entropy decrease by $k \ln 2$.

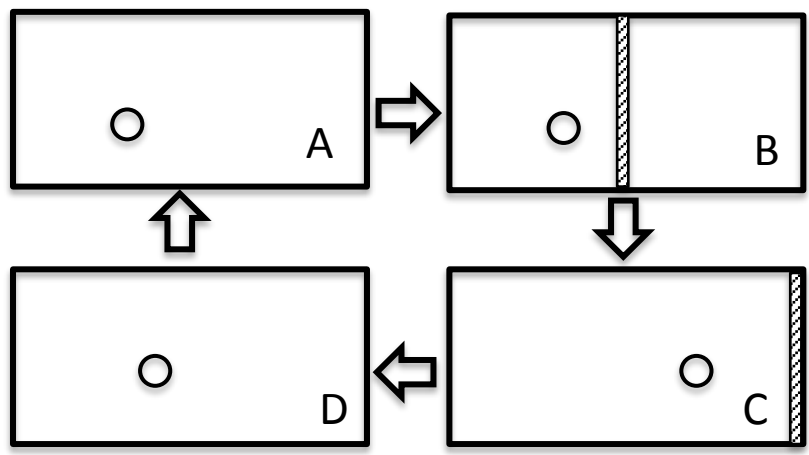

**FIG. 1** Schematic of the original Szilard engine. (A) There is a single molecule inside the box. (B) Inserting a piston divides the box into two halves with the molecule trapped on one side of the piston. (C) The pressure from the single molecule pushes the piston to one end of the box performing work. (D) Finally, the piston is removed and the system completes its cycle.

Because the second law of thermodynamics does not allow entropy decrease of the universe, Szilard concluded that there must be an entropy increase somewhere else by at least $k\ln 2$ during the operation of the engine by stating that "The amount of entropy generated by the measurement (of the location of the molecule) may, of course, always be greater than this fundamental amount, but not smaller" [10]. Later, Landauer and Bennett argued that measurement of the molecular position could be performed without costing any energy, and hence assigned the $k\ln 2$ entropy to memory erasure [11, 12]. Although Szilard and Bennett assigned the required entropy change of $k\ln 2$ to different stages of the engine cycle, they both agree on one point that the isothermal expansion, from B to C in Fig. 1, should proceed spontaneously. In this Letter, I will show that when the piston is properly considered as under thermal contact with the surrounding, it is not possible for the Szilard engine to proceed from step B to step C, Figure 1. The breakdown of the Szilard engine at step B has a significant consequence: the reservoir does not need to transfer $kT\ln 2$ of heat to the molecule and hence suffers no entropy decrease.

The initial concept of the Szilard engine assumed the piston as a passive object of mass M such as that appears in classical mechanics. The internal energy contained within the piston was not taking into account. The initial momentum of the piston is conveniently assumed to be zero and the piston gains momentum only as a result of collision with the molecule. The treatment of the piston as a passive object is fundamentally flawed. Because the piston is in thermal equilibrium with its surroundings, thermal energy associated with the mass of the piston must be taken into consideration. There are two components of the thermal energy associated with the piston. The first arises from the thermal energy of all atoms within the piston. At a sufficiently high temperature, room temperature for example, the energy contained in the piston equals $3NkT$ where $N$ is the number of atoms inside the piston and $k$ the Boltzmann constant. Connecting to this thermal energy, there is also an energy fluctuation of $\sqrt{3N}kT$. The second energy component has its origin in the Brownian motion of the piston as a whole and it amounts to energy $kT$. The average energy of the single molecule is $\frac{3}{2}kT$ which is subject to fluctuations at the level ~ $kT$.

Consider a collision between the single molecule and the piston, the maximum amount of energy can be transferred from the molecule to the piston in a single collision is in the order of $kT$. This is a very small amount of energy. Never the less, one can argue that no matter how small it is, this energy is useful for doing work. This is the foundation of the Szilard engine. However, one must not forget that the energy within the piston fluctuates by $\sqrt{3NkT}$ which is significantly higher than $kT$ for a realistic piston. For $N$ to be the order of $10^{23}$, the energy fluctuation within the piston is ~$10^{12}$ orders of magnitude greater than $kT$. I will discuss the case where $N$~ 1 later. The energy within the piston fluctuates by $\sqrt{3NkT}$ regardless whither there is a single molecule in the box or not, or whither the molecule is to the left or right of the piston. This is a direct consequence that the piston is under equilibrium at temperature T. The deposition of energy $kT$ to a system that is subject to energy fluctuation of $\sqrt{3NkT}$ has no effect on the energy of the piston. This is analogous to the situation of the water level in a lake. The water level fluctuates by ~ mm due to water flowing in and out of the lake, and hence adding a single raindrop to the lake should have effectively zero influence.

One also needs to consider the fact that the molecule does not collide with a featureless slab of solid. The piston is composed of atoms and the collision takes place between the molecule and individual atoms on the surface of the piston, Fig. 2. Each atom has an average energy of 3 $kT$ which itself fluctuates by ~ $kT$. The fact that the energy of individual atoms within the piston fluctuates has an important consequence: for an arbitrary collision between the molecule and the piston, ~$kT$ energy can be transferred to the piston or from the piston to the molecule. Statistically, the net energy transfer should be zero after a large number of collisions have taken place. When the molecule approaches the piston, its collision with a specific surface atom is controlled by their relative energies and momenta. Under situations where energy is transferred from the piston to the molecule, the molecule can be regarded as have been compressed by the piston. The only possibility for continuous energy transfer from the surrounding to the piston via the single molecular expansion is that you treat the piston as a mass without internal structure. Even so, the Brownian motion of such a mass would prevent the expansion of the single molecule.

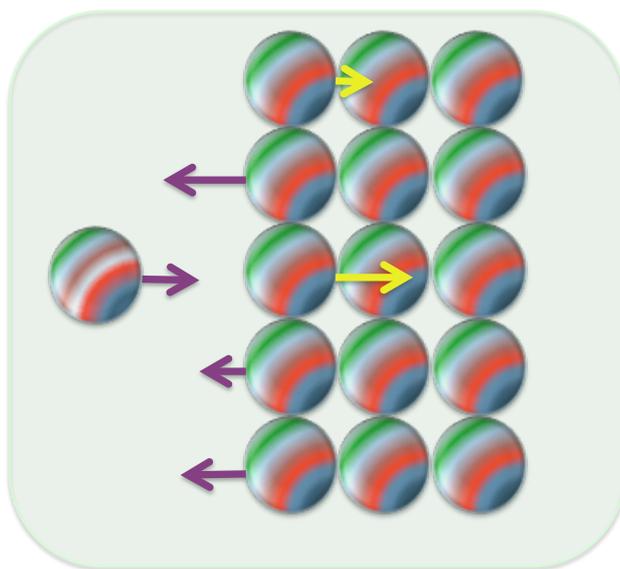

**FIG. 2** Schematic showing the interaction between a single molecule and atoms within the piston. Arrows indicate the velocities of molecule/atoms. The collision between the single molecule and an atom can result in energy transfer in either direction. Statistically, there is equal probability that the piston is pushed by the molecule as that the molecule compressed by the piston.

We can reduce $\sqrt{3NkT}$ by using lighter pistons. For example, a nano-piston formed by just one single, planar, molecule. In this case the energy fluctuation of the piston is reduced to $\sim kT$. Such a nano-piston would move between the two ends of the box in a random fashion driven by its thermal energy. It is impossible for the piston to be pushed to one end of the box by the single molecule. The same can be said about the molecule which cannot be squeezed by the piston to one end of the box. The situation is equivalent to two independent molecules moving randomly in the box. A similar conclusion has been reached by Norton in his analysis of the expansion and compression of a single molecule [13].

The single molecule Szilard engine is thus unable to do work. How about two or more molecules? The answer is that you need many many more than just a few molecules. The energy transferred from collisions of all molecules with the piston must exceed the energy fluctuation of $\sqrt{3NkT}$. This condition is not usually discussed in typical thermal expansion processes because in most cases the condition is satisfied. If we use a piston that contains $10^4$ atoms, we require about 100 molecules on one side of the piston in the box to operate the engine. With 100 molecules, the probability of keeping them all on the same side of the piston without relying on isothermal compression becomes diminishingly small. Therefore, we face an un-resolvable dilemma. If the number of molecules is small, the isothermal expansion step cannot proceed; if the number of molecules is large, there is no way to push them to one half of the box without doing necessary work.

In summary, the classical Szilard single molecule engine does not work. The piston will not be displaced by the pressure from a single molecule. There is no need to search for the $k\ln 2$ entropy to compensate for the apparent entropy decrease during the isothermal expansion, because the expansion simply cannot take place. The critical point of confusion in the original concept of the Szilard engine is treating the piston as a mass without internal atomic structure as well as ignoring the fluctuation of its internal energy.

# References

[1] A. Berut, A. Arakelyan, A. Petrosyan, S. Ciliberto, R. Dillennschneider, and E. Lutz, Nature (London) **483**, 187, (2012).

[2] E. Roldan, I. A. Martinez, J. M. R. Parrondo, and D. Petrov, Nat. Phys. **10**, 457 (2014).

[3] D. Mandal, and C. Jarzynski, PNAS **109**, 11641 (2012).

[4] J. V. Koski, V. F. Maisi, J. P. Pekola, and D. V. Averin, PNAS **111**, 13786 (2014).

[5] S. Toyabe, T. Sagawa, M. Ueda, E. Muneyuki, and M. Sano, Nat. Phys. **6**, 988 (2010).

[6] P. Strasberg, G. Schaller, T. Brandes, and M. Esposito, Phys. Rev. Lett. **110**, 040601 (2013).

[7] A. C. Barato, and U. Seifert, Phys. Rev. Lett. **112**, 090601 (2014).

[8] S. Deffner, and C. Jarzynski, Phy. Rev. X. **3**, 041003 (2013).

[9] S. W. Kim, T. Sagawa, S. De Liberato, and M. Ueda, Phys. Rev. Lett. **106**, 070401 (2011).

[10] L. Szilard, Zeitschrift fur Physik. **53**, 840 (1929).

[11] R. Landauer, IBM J. Res. Dev. **5**, 183 (1961).

[12] C. H. Bennett, Int. J. Theor. Phys. **21**, 905 (1982).

[13] J. D. Norton, Entropy **15**, 4432 (2013).